# MOEVC: A MIXTURE-OF-EXPERTS VOICE CONVERSION SYSTEM WITH SPARSE GATING MECHANISM FOR ACCELERATING ONLINE COMPUTATION

*Yu-Tao Chang[1], Yuan-Hong Yang[1], Yu-Huai Peng[3], Syu-Siang Wang [2], Tai-Shih Chi [1], Yu Tsao [2], Hsin-Min Wang[3]*

[1] Department of Electrical and Computer Engineering, National Chiao Tung University, Hsinchu, R.O.C
[2] Research Center for Information Technology Innovation, Academia Sinica, Taipei, Taiwan
[3] Institute of Information Science, Academia Sinica, Taipei, R.O.C

## ABSTRACT

With the recent advancements of deep learning technologies, the performance of voice conversion (VC) in terms of quality and similarity has been significantly improved. However, heavy computations are generally required for deep-learning-based VC systems, which can cause notable latency and thus confine their deployments in real-world applications. Therefore, increasing the online computation efficiency has become an important task. In this study, we propose a novel mixture-of-experts (MoE) based VC system. The MoE model uses a gating mechanism to specify optimal weights to feature maps to increase the VC performance. In addition, assigning sparse constraints on the gating mechanism can accelerate online computation by skipping convolution process by zeroing out redundant feature maps. Experimental results show that by specifying suitable sparse constraints, we can effectively increase the online computation efficiency with a notable 70% FLOPs (floating point operations per second) reduction while improving the VC performance in both objective evaluations and human listening tests.

## 1. INTRODUCTION

Voice conversion (VC) aims to convert speech signals from one speaker to another without changing the linguistic content. There are a wide variety of VC applications, such as personalized text-to-speech system (TTS), speech to speech translation, speaking and hearing-aid devices, and singing voice conversion [1]. Numerous models have been proposed as the fundamental tool for the VC task. Well-known examples include Gaussian mixture model (GMM) [2, 3], exemplar-based models [4], variational autoencoder (VAE) [5, 6], and generative adversarial network (GAN) [7]. Traditionally, VC is performed in a frame-to-frame conversion manner. A notable limitation is that temporal information of the voice signals cannot be well characterized. More recently, long short-term memory (LSTM), which shows good capability of modeling temporal information, has been proposed for the VC task and achieved satisfactory performance [8, 9, 35]. In the meanwhile, fully convolutional network (FCN) model has also been used to perform sequence-to-sequence conversion and obtained higher mean opinion score (MOS) scores as compared to frame-to-frame-based conversion methods [10].

To develop VC on real-world applications, there are generally two requirements: (1) using a lite model that requires small storage; (2) being able to generate converted voice with minimal computational cost and latency. For classification tasks, many algorithms have been proposed to address these two requirements. However, for regression tasks (such as VC), there are only few studies in the literature. In [11] and [12], the authors proposed to use parameter quantitation and scaling techniques to compresses the deep-learning models for speech enhancement (SE) tasks. The results demonstrate that when adequately design the compression rate, we can strike a good balance between the model size and SE performance. In this study, we focus on the second requirement that aims to reduce the online computation cost and propose a novel mixture-of-experts voice conversion (MoEVC) system with sparse constraints on the gating mechanism.

The authors of [13] extended the conventional MoE model [14] to DeepMoEs, an architecture which is capable of reducing computational complexity of a deep-learning based models with convolution layers. The model combines a base convolutional network with an embedding network along with a sparse gating network [13]. The input passes through both embedding network and sparse gating network, and the resulting gating values is used to select the channels in each layer. The model achieves comparable or even better performance as compared to other state-of-the-art channel pruning methods [15,16,17,18]. We believe that the DeepMoEs architecture can be suitably used in the VC system to reduce the online computation cost without compromising much speech quality.

The proposed MoEVC uses the FCN-VAE [19] as the base architecture and integrates the MoE module to increase the online computation efficiency. Although the DeepMoEs model has the potential to be applied to any architecture, three issues confine a direct combination of DeepMoEs and FCN-VAE. First, the DeepMoEs model was proposed for classification of image with a fixed size, while FCN-VAE supports speech conversion with arbitrary speech lengths. Second. all gating values in DeepMoes are controlled by a single input; however, FCN-VAE accommodates an additional speaker code as input. Third, DeepMoEs takes a single image as the input while the two inputs to the FCN-VAE model, namely spectral features and speaker code, are in different formats.

In this paper, we design an auxiliary embedding network with different structures for both encoder and decoder in the FCN-VAE model in order to effectively combine DeepMoEs with FCN-VAE. We investigate the relationship between the reduction in FLOP (floating point operations per second) and the quality of converted speech. To alleviate extensive time-consuming listening tests, we first evaluate the VC results yield by various sparse constraints (computation acceleration rates) using the MOSNet [20], which serves as a learning-based objective evaluator. We also validate the consistency of the listening test results and the MOSNet score. Experimental results show that the proposed MoEVC system can reduce FLOPs by a significant 70%, while even increasing the speech quality as compared to the original FCN-VAE system.

The remainder of this paper is organized as follows. Section 2 reviews related works. Section 3 presents the proposed MoEVC system. Section 4 presents the experimental setup and results. Finally, Section 5 provides the conclusion remarks of this study.

## 2. RELATED WORKS

In this section, we present related works to the proposed MoEVC.

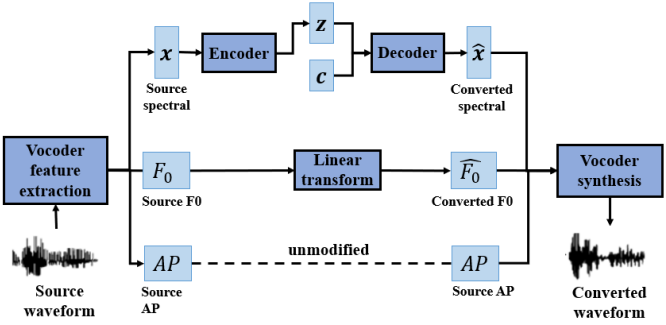

Fig. 1 The architecture of the VAE-VC framework

## 2.1 VAE-VC

Fig. 1 illustrates the architecture of the VAE-VC system [5]. The VAE-VC is formed by an encoder-decoder architecture During training, given an input spectral frame $x$, the encoder $E_\theta$, with the parameter set $\theta$, encodes $x$ into a latent code: $z = E_\theta(x)$. The speaker code, $c$, of the input frame, along with $z$, are processed by the decoder $G_\Phi$ with parameter set $\Phi$ to reconstruct the input. The reconstruction process can be written as:

$$\bar{x} = G_\Phi(z, c) = G_\Phi(E_\theta(x), c). \quad (1)$$

The model parameters can be obtained by maximizing the variational lower bound:

$$L_{vae}(\theta, \Phi; x, c) = L_{recon}(x, c) + L_{lat}(x), \quad (2)$$

where $L_{recon}$ denotes a reconstruction term, and $L_{lat}$ regularizes the encoder to align the approximate posterior with the prior distribution.

During the conversion phase, given the source speech features, $x$, we can then formulate the conversion $f(.)$ and accordginly obtain the converted $\hat{x}$ via:

$$\hat{x} = f(x, \hat{c}) = G_\Phi(z, \hat{c}) = G_\Phi(E_\theta(x), \hat{c}). \quad (3)$$

Then, $\hat{x}$ is fed into a vocoder to obtain converted speech waveforms.

So far, many extensions have been proposed based on the VAE-VC framework. In [7], the GAN model was integrated with VAE model. An auxiliary classifier was also adopted to enhance disentanglements to attain further improvement [21]. [25] shows that an autoencoder with carefully designed bottleneck can perform VC task. Later on, Wavent-based models [32, 33, 34] are used to increase speech quality in terms of naturalness and similarity.

## 2.2 ACVAE-VC

In [19], Kameoka et al. proposed an auxiliary classifier variational autoencoder (ACVAE) VC framework; Fig. 2 shows the overall arachitecture of the ACVAE-VC system. The ACVAE-VC is based on an vallina VAE-VC, where gated CNNs were used for both encoder and decoder. For the $l$-th input, with having height and width of $Q_l$ and $N_l$, and with a speaker code $c$, which is a one-hot vector with a length equaling to the number of speakers used for training, the coresponded input-output relation in this layer can be formulated as:

$$h'_{l-1} = [h_{l-1}; c_{l-1}], \quad (4)$$

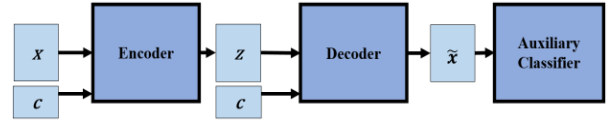

Fig. 2 The overall architecture of the ACVAE-VC system.

$$h_l = (W_l * h'_{l-1} + b_l) \odot \sigma(V_l * h'_{l-1} + d'_l), \quad (5)$$

where $h_{l-1}$ and $h_l$ represent input and ouput of the $l$-th hidden layer, respectively; $c_{l-1}$ is a 3D array consisting of a $Q_{l-1}$-by-$N_{l-1}$ tiling of copies of $c$, $\{W_l, V_l\}$ and $\{b_l, d_l\}$ are the weights and biases of the convolutional layer; $[h_l^T, c_l^T]^T$ is the concatenation of $h_l$ and $c_l$ along the channel dimension; $\odot$ and $\sigma$ denote the elementwise multiplication and the elementwise sigmoid function, respectively.

As shown in Fig. 2, the ACVAE-VC system uses an auxiliary classifier to encourage model to generate output with considering the speaker code. Specifically, the classifier approximates posteriori probability of speaker $p(c|x)$ and is trained to maximize the lower bound of mutual information between decoder output and speaker code. As derived in the [19], the objective function becomes

$$J(\Phi, \theta) + \lambda_L \mathrm{MI}(\Phi, \theta, \psi) + \lambda_I \mathrm{CE}(\psi), \quad (6)$$

where $J$ is the variational lower bound, MI is the mutual information lower bound, and CE is the cross-entropy of the auxiliary classifier. $\Phi, \theta$, and $\psi$ denote the parameters for the encoder, decoder, and auxiliary classifier, respectively.

## 2.3 Deep mixture of experts (DeepMoEs)

Motivated by the work of [22] that demonstrated the advantages of stacking two layers of MoEs, Wang et al. proposed the DeepMoEs architecture for image classification task, which is capable of increasing computational efficiency and has the potential to be applied to any model that has convolutional layers [13]. Specifically, through the sparse gating mechanism, DeepMoEs allows each convolutional layer to dynamically select only part of output feature maps to be activated at inference time. Fig. 3 shows the architecture of the DeepMoEs, which consists of three basic components: (1) Base convolutional network; (2) Embedding network(EMN); (3) Sparse gating network (SGN). The base convolutional network and embedding network share the same input, where the embedding network maps input to embedding $e$. Then the embedding is transformed by the SGN into a gating vector:

$$g = G^l(e) = \mathrm{ReLU}(W_g^l \cdot e), l = 1 \dots, L, \quad (7)$$

where $g$ is the gating vector, $L$ is the total number of convolutional layer in the base convolutional network. As a result, DeepMoEs scales each feature map by the gating values generated from SGN. For a convolutional layer with tensor input $h$ having spatial-resolution $W_{in} \times H_{in}$ and $C_{in}$ input channels, a $C_{in} \times k \times k \times C^{out}$ convolutional kernel $K$, and a set of gating values, $= [g_1, g_2 \dots g_{C_{in}}]^T$ generated by SGN, the output $y$ can be calculated by:

$$y = \sum_{i=1}^{C_{in}} g_i K_i * h_i = \sum_{i=1}^{C_{in}} g_i F_i(h), \quad (8)$$

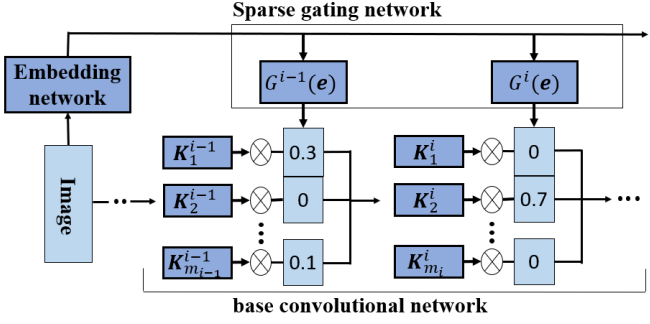

Fig. 3 The architecture of the DeepMoEs model.

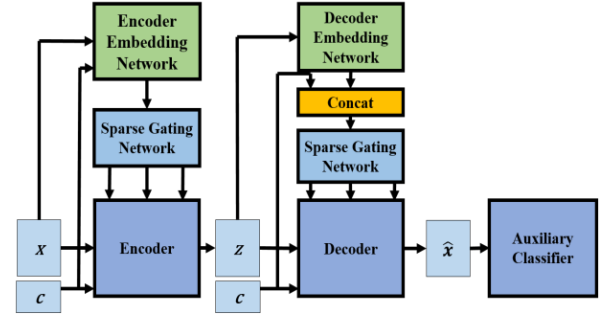

Fig. 4 The architecture of the proposed MoEVC system.

where $F_i(\mathbf{h})$ denotes a function of $\mathbf{h}$.

Because the gating values are generated by the ReLU activation function, some elements may be zeros by introducing the sparse constraint. To further control the sparsity of SGN and the diversity of embedding network, the authors proposed to introduce the $L1$ regularization term on the SGN outputs, along with the auxiliary classification loss. Accordingly, the overall loss function becomes

$$L_{all}(\mathbf{s}; \mathbf{t}) = L_b(\mathbf{s}; \mathbf{t}) + \lambda L_g(\mathbf{s}) + \mu L_e(\mathbf{s}; \mathbf{t}), \qquad (9)$$

where $\mathbf{s}$ and $\mathbf{t}$ are the input image and the corresponding label, $L_b$ is the loss from the base convolutional network, $L_g$ is the $L1$ penalty term, and $L_e$ is the auxiliary classification loss; $\lambda$ and $\mu$ are coefficients to determine the weights between model accuracy, sparsity, and embedding diversity.

### 2.4 MOSNET: A learning based objective evaluator

An efficient and effective quality measurement of generated speech has been a longstanding problem in TTS, SE, and VC tasks. In the VC community, objective measures, such as the Mel-cepstral distance (MCD) [23] and global variance (GV) [24], are widely used for automatically measuring the quality of converted speech. As reported in many previous studies [3,25], such metrics are not always correlated with human perception perfectly as they mainly measure the distortions of acoustic features. Subjective listening tests, such as the MOS and similarity scores, could represent more realistically the intrinsic naturalness and similarity of a VC system. However, these human-involved evaluations are usually time-consuming and expensive as they need a large number of participants to conduct listening tests and provide perceptual ratings. Recently, Lo et al. proposed a MOSNet, which uses a deep learning based model to perform speech naturalness and similarity assessment [20]. The results showed that the predicted MOS and similarity scores are well correlated with human similarity ratings. In this paper, we adopt MOSNet as a learning-based objective evaluator (formed by a CNN-BLSTM architecture) to evaluate the VC performance.

### 3. THE PROPOSED MOEVC SYSTEM

In order to combine the DeepMoEs and FCN-VAE VC systems, we mainly focus on the design of the embedding network to let the model meet the following requirements: (1) The decoder is not affected by the speaker code during the conversion phase; (2) The embedding network can process speech signals with any specific lengths; (3) The embedding network can combine inputs from different domains. The proposed MoEVC system is shown in Fig. 4.

Different from the original DeepMoEs model that only uses one embedding network to control all gating values in the base network, the proposed MoEVC adopts two embedding networks to control the gating values in the encoder and decoder separately; these two embedding networks are termed encoder/decoder embedding network (EEN/DEN), as shown in Fig. 4. To fulfil the above-mentioned three requirements, first, we design the DEN to accommodate two inputs: the latent code and the speaker code. In this way, the decoder embedding is independent of source speaker code during the conversion phase. This is to meet the requirement of (1).

To meet the requirements (2) and (3), we used the convolutional layers of the EEN because the CNN structure has been proven to be an effective feature extractor in previous works [26,27]. Moreover, we used the temporal pooling layer that is proposed by [28] to average the input along the time dimension so that we can get a fix-sized vector; then the final embedding is generated by several fully-connected layers for non-linear transformations. As for the input of the EEN, we first create a tensor which is a $Q$-by-$N$ tiles of $\mathbf{c}$, where $Q$ and $N$ are spatial-resolution of $\mathbf{x}$. For the architecture of DEN, we combine a sequential-autoencoder [29] with several fully-connected non-linear transformations. Specifically, we jointly train a sequential-autoencoder to reconstruct $\mathbf{z}$, and use the encoder part of the sequential-autoencoder to perform feature extraction. The feature is concatenated with $\mathbf{c}$ and passed through fully-connected non-linear transformations to generate the final embedding.

Finally, to train the model, the overall loss function becomes:

$$L_{all}(\mathbf{x}; \mathbf{c}) = L_{base}(\mathbf{x}; \mathbf{c}) + \alpha L_{ae}(\mathbf{z}) + \beta L_{spc}(\mathbf{x}), \qquad (10)$$

where $L_{base}$, $L_{ae}$, and $L_{spc}$ are the losses of the base convolutional model, DEN sequential-autoencoder, $L1$ norm of SGN, respectively.

### 4. EXPERIMENT

#### 4.1 Experiment setup

We evaluated the proposed MoEVC system on the Voice Conversion Challenge 2016 dataset. The speech signals were recorded by US English speakers in a professional recording studio without noticeable noise effects. The dataset consisted of 162 utterances for training and 54 utterances for evaluation from each of 5 source and 5 target speakers. We used a subset of speakers for training and evaluation. Specifically, we used all training and evaluation data provided by two female speakers 'SF2' and 'SF3', and two male speakers 'SM1' and 'SM2'. Thus, there were totally four types of conversions, namely female to female (F2F), male to male (M2M), female to male (F2M), and male to female (M2F). We report the average scores for each conversion type.

All the speech signals were sampled at 16KHz, and the WORLD vocoder [30] was used to extract 513 dimensional *SPs*, 513-dimensional aperiodic components (*Aps*), and fundamental frequency ($F_0$) at every 5ms; 36 dimensional-MCCs were further extracted from *SPs* and further standardized per pixel dimension based on the corresponding mean and the standard deviation; these two statistics were saved and used in the conversion phase.

MoEVC used ACVAE as the base convolutional model, which was trained using the same architecture as in [19]. Additionally, EEN, DEN, and SGN were added on top of the base convolutional model. We adjusted the number of CNN kernels and fully connected units in the EEN and DEN so that the total parameters were marginally increased while the total FLOPs were increased by less than 1% (before enforcing sparse constraints). The notable difference results of the increase of parameters and the increase of FLOPs is due to the fact that CNN layer has few parameters but requires high FLOPs, and the increase of parameters is mainly due to the increase of fully connected units which require few FLOPs for inference.

During the training phase, we randomly sampled 512 frames with overlap from each utterance, and all models were trained with 5000 epochs using the Adam [31] optimizer with initial learning rate, $B_1$, and $B_2$ set to 0.001, 0.9, 0.999, and the batch size was 64.

### 4.2 Experiment Results

In this study, we focused our attention on the comparison of online computation cost and the naturalness of converted speech. We first used the MOSNet to evaluate the converted utterances with different sparse constraints. When setting a larger sparse constraint value, the gating values will be sparser, more feature maps will be zeroed out, and finally the online computation cost can be reduced. Since we have investigated a wide variety of spare constraint values, it is difficult to conduct listening tests for all the setups. Therefore, we decided to first apply the MOSNet to objectively evaluate the converted speech. Then we conducted listening tests to confirm the trends of the predicted MOSNet scores with the listening results.

*4.2.1 MOSNet results*

Fig. 5 shows the correlation of computational cost in terms of FLOP reduction rates (FRR) and MOSNet scores. Please note that we intend to unbiasedly observe the relations between FRRs and MOSNet scores and thus did not use any post-filtering methods, such as GV [24] and maximum likelihood parameter generation (MLPG). Meanwhile, $F_0$ was converted by a linear mean-variance transformation in the log domain, and *APs* were kept unmodified.

From Fig. 5, we first note that when we slightly increased the sparse constraints to enable a higher FRRs (from points Ⓐ to Ⓑ), the MOSNet scores are actually increased. Next, the scores of the MOSNet reached to the highest point in Ⓑ, where the FRRs and MOSNet scores were 72% and 3.2, respectively. When further increasing the FRR, the MOSNet scores started to decrease (as can be seen from the decrease from Ⓑ to Ⓒ). This trend is often observed in model compression and acceleration tasks: With an adequate compression with removal of redundant components, deep-learning models can yield better performance; further compressions may result in performance degradations.

*4.2.2 Human listening test results*

Next, we verified the findings from the MOSNet evaluations by conducting human listening tests. The listening test was conducted by a total of 11 subjects to listen to the converted speech of Ⓐ, Ⓑ, and Ⓒ in Fig. 5 for the MoEVC system, and also the converted speech given by the original ACVAE. Each subject was asked to score the naturalness and the similarity between the original speaker of 12 utterances, which were also evaluated by MOSNet. Tables 1 and 2 show the MOS scores of naturalness and the similarity scores, both are in the score scale of 1 to 5, the higher the better.

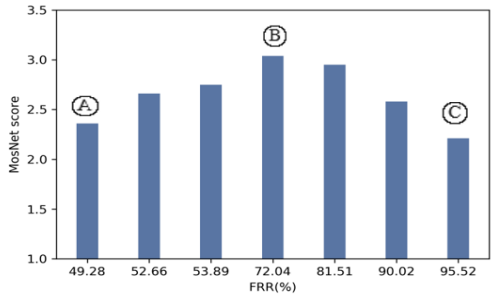

Fig. 5 MOSNet result of ACVAE generating method

From both tables, we note that both the naturalness and similarity results are quite consistent to the ones given by the MOSNet as shown in Fig. 5. The results of Ⓑ are better than Ⓐ and Ⓒ. Moreover, Ⓑ can yield even better naturalness and similarity scores than the conventional ACVAE-VC (where no sparse constraint was applied and thus no acceleration was achieved).

Table 1. Naturalness scores on the converted speech produced by MoEVC with different FRRs and ACVAE.

|  | FRR | F2F | F2M | M2F | M2M | AVE |
|---|---|---|---|---|---|---|
| Ⓐ | 49% | 3.00 | 2.00 | 1.88 | 1.25 | 2.00 |
| Ⓑ | 72% | **4.05** | **3.08** | **3.50** | **4.05** | **3.54** |
| Ⓒ | 95% | 2.25 | 1.69 | 1.69 | 2.88 | 1.98 |
| ACVAE | 0% | 3.70 | 2.63 | 3.23 | 3.60 | 3.17 |

Table 2. Similarity scores on the converted speech produced by MoEVC with different FRRs and ACVAE.

|  | FRR | F2F | F2M | M2F | M2M | AVE |
|---|---|---|---|---|---|---|
| Ⓐ | 49% | 3.00 | 1.88 | 1.69 | 1.50 | 1.94 |
| Ⓑ | 72% | **3.70** | **2.28** | 2.20 | **3.60** | **2.71** |
| Ⓒ | 95% | 2.38 | 1.75 | 1.50 | 3.00 | 1.98 |
| ACVAE | 0% | 3.25 | 2.18 | **2.23** | 3.60 | 2.61 |

## 5. CONCLUSION

The main contribution of this study is twofold. First, we confirmed the effectiveness of introducing the DeepMoEs model to accelerate the online computation for the VC task. Based on our experimental results, the proposed MoEVC system can reduce more than 70% of FLOPs without harming and even increasing the quality of converted speech in terms of both naturalness and similarity of converted speech. Second, we present that the MOSNet can be used as an effective learning-based objective evaluator for the VC task. Because it was prohibitive to conduct extensive human listening tests, we decided to use MOSNet to predict MOS scores. We further confirmed that the predicted scores are consistent to the results of human listening tests. Hopefully, the findings of this study can promote the research of model compression and online computation acceleration for VC. In the future, we will test the compatibility of the MoEVC with advanced vocoder systems and learning algorithms.